# Complexity and Applications of Edge-Induced Vertex-Cuts

Marko Samer[1,*] and Stefan Szeider[2,**]

[1] Institute of Information Systems (DBAI)
Vienna University of Technology, Austria
samer@dbai.tuwien.ac.at
[2] Department of Computer Science
Durham University, UK
stefan.szeider@durham.ac.uk

**Abstract.** Motivated by hypergraph decomposition algorithms, we introduce the notion of *edge-induced vertex-cuts* and compare it with the well-known notions of edge-cuts and vertex-cuts. We investigate the complexity of computing minimum edge-induced vertex-cuts and demonstrate the usefulness of our notion by applications in network reliability and constraint satisfaction.

## 1 Introduction

One of the basic concepts in graph theory is connectivity: the minimal number of edges or vertices that disconnect a graph when removed. Such sets of edges or vertices that disconnect a graph are called *cuts*. Cuts that are minimal with respect to some measure are called *minimum cuts* or short *mincuts*. Mincuts have many important applications in network theory and combinatorial optimization. For example, consider a graph that models a telecommunication network, i.e., each vertex represents a communication station and each edge represents a communication line connecting two communication stations. In order to check network reliability, there are now two classical approaches:

The first one is to determine the minimal number of communication lines that have to fail (e.g., by cutting the lines) in order to disconnect the network. This approach corresponds to computing a minimum *edge-cut* in the graph. The second one is to determine the minimal number of communication stations that have to fail (e.g., by power failure or demolition) in order to disconnect the network. This approach corresponds to computing a minimum *vertex-cut* in the graph.

However, there is a third possibility which corresponds to our new *edge-induced vertex-cut*: Determine the minimal number of communication lines that have to be misused (e.g., by sending a computer virus or high voltage) in order to shut down all directly connected communication stations which thereby disconnect the network.

To illustrate the difference between these three kinds of minimum cuts, consider the graph in Figure 1. An example of a minimum *edge-cut* separating $s$ and $t$ in this graph is given by $\{I, J, O, P\}$ and therefore of size $4$. Moreover, a minimum *vertex-cut* separating $s$ and $t$ is given by $\{e, h, i\}$ and therefore of size 3. Finally, a minimum *edge-induced vertex-cut* separating $s$ and $t$ is given by $\{E, F\}$ and therefore of size 2. It is a well-known fact that minimum vertex-cuts can never be larger than minimum edge-cuts in ordinary graphs. In the more general setting of hypergraphs, however, this is not the case. On the contrary, even in hypergraphs minimum edge-induced vertex-cuts can never be larger than edge-cuts or vertex-cuts as we will show in this paper.

---

[*] Research supported by the Austrian Science Fund (FWF), P17222-N04.
[**] Research supported in part by the Nuffield Foundation, NAL/01012/G.

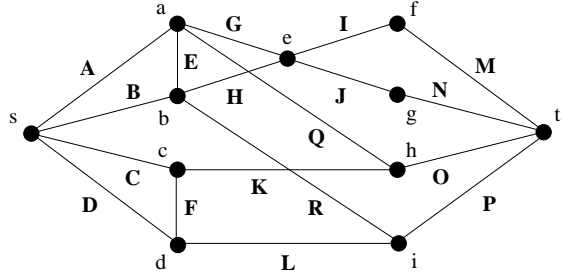

**Fig. 1.** Graph example to illustrate mincuts

In addition to network reliability, several other applications of edge-induced vertex-cuts in the area of combinatorial optimization are conceivable. Our original motivation for investigating edge-induced vertex-cuts, however, comes from the area of constraint satisfaction; in particular, from hypertree decomposition [7].

In hypertree decomposition, a hypergraph is transformed into clusters of hyperedges that are organized as a tree which has to satisfy several conditions. The connection to our work here is that each node of the resulting tree represents a (not necessarily minimal) edge-induced vertex-cut in the underlying hypergraph, i.e., the vertices are used to disconnect the hypergraph but the hyperedges are used to measure the hypertree-width. Thus, a thorough understanding of edge-induced vertex-cuts may also give a better understanding of hypertree decompositions and related methods. Moreover, our results in this paper show from a theoretical point of view that graph partitioning heuristics are in principle unsuitable for decomposing hypergraphs into hypertrees since graph partitioning heuristics aim at minimizing edge-cuts (resp. vertex-cuts) while hypertree decompositions aim at minimizing edge-induced vertex-cuts.

This paper is organized as follows: In Section 2, we introduce the basic terms used in the remainder of this paper. Then, in Section 3, we recall the definitions of edge-cuts and vertex-cuts and give several examples. Afterwards, in Section 4, we define our new edge-induced vertex-cut and prove that the size of a minimum edge-induced vertex-cut in a hypergraph is always smaller than or equal to the size of a minimum edge-cut and a minimum vertex-cut. In Section 5, we compare the time complexity of the different cut variants; in particular, we prove that computing an edge-induced vertex-cut of minimal size is NP-hard. Finally, we describe important applications of edge-induced vertex-cuts in Section 6 and conclude in Section 7.

## 2 Preliminaries

A *hypergraph* is a tuple $(V, E)$ of a non-empty set $V$ of vertices and a set $E \subseteq 2^V \setminus \{\emptyset\}$ of (hyper)edges. A *graph* is a hypergraph where each edge $e \in E$ contains exactly two vertices, i.e., $|e| = 2$ for all $e \in E$. A *path* in a hypergraph $H = (V, E)$ is a sequence $v_1, v_2, \ldots, v_k$ of vertices in $V$ where $v_i \neq v_j$ for all $1 \leq i < j \leq k$ and for each vertex $v_i$ with $1 \leq i < k$ there exists $e \in E$ such that $v_i, v_{i+1} \in e$. We say two vertices $v$ and $w$ are connected in $H$ if there exists a path of the form $v, \ldots, w$ in $H$. Finally, we say a hypergraph is connected if any two of its vertices are connected.

## 3 Edge-Cuts and Vertex-Cuts

In this section, we will formally define what we understand by edge-cuts and vertex-cuts as well as by their restricted variants $s$-$t$-edge-cuts and $s$-$t$-vertex-cuts. We will exemplify edge-cuts and vertex-cuts by finding respective minimum cuts in the hypergraph shown in Fig. 2 and the graph shown in Fig. 3. Note that these examples will also demonstrate the difference of edge-cuts and vertex-cuts to our new edge-induced vertex-cuts, which we will introduce in Section 4.

The following definitions are straight-forward generalizations of "separating sets" of graphs defined in [2]. Let us start with *edge-cuts*:

**Definition 1 (Edge-Cut).** *Let $H = (V, E)$ be a hypergraph and $s, t \in V$ with $s \neq t$. A set $C \subseteq E$ is an $s$-$t$-edge-cut in $H$ if $s$ and $t$ are not connected in $H' = (V, E \setminus C)$. A set $C \subseteq E$ is an edge-cut in $H$ if there are two distinct vertices $v, w \in V$ such that $C$ is a $v$-$w$-edge-cut in $H$. The size of an edge-cut $C$ is its cardinality. A minimum edge-cut is an edge-cut of minimal size, and a minimum $s$-$t$-edge-cut is an $s$-$t$-edge-cut of minimal size. The edge-connectivity $\lambda(H)$ of $H$ is the size of a minimum edge-cut in $H$, and the $s$-$t$-edge-connectivity $\lambda_H(s, t)$ of $H$ is the size of a minimum $s$-$t$-edge-cut in $H$.*

*Remark 1.* In the literature, *edge-cuts* are often denoted as *cuts* and defined in a slightly modified way [2]: $C \subseteq E$ is a *cut* in $G$ if there exists a non-empty subset $V' \subset V$ such that $C = \{e \in E \mid e \cap V' \neq \emptyset, \ e \cap (V \setminus V') \neq \emptyset\}$. It is easy to see that our definition in Definition 1 is more general than this one. However, when considering *minimum* edge-cuts only, both definitions are equivalent.

*Example 1.* A minimum $s$-$t$-edge-cut in the hypergraph in Fig. 2 is given by $\{I, L, M\}$, which separates vertex $s$ from vertex $t$. Similarly, a minimum $s$-$t$-edge-cut in the graph in Fig. 3 is given by $\{R, S, Y, Z\}$, which also separates vertex $s$ from vertex $t$.

A minimum edge-cut in the hypergraph in Fig. 2 is given by $\{P, R\}$, which separates vertex $j$ from all other vertices in the hypergraph. Similarly, a minimum edge-cut in the graph in Fig. 3 is given by $\{O, W, X\}$, which separates vertex $m$ from all other vertices in the graph.

Let us now define *vertex-cuts* in analogy to edge-cuts:

**Definition 2 (Vertex-Cut).** *Let $H = (V, E)$ be a hypergraph and $s, t \in V$ with $s \neq t$. A set $C \subseteq V \setminus \{s, t\}$ is an $s$-$t$-vertex-cut in $H$ if $s$ and $t$ are not connected in $H' = (V \setminus$* $C, E$). *A set $C \subseteq V$ is a* vertex-cut *in $H$ if there are two distinct vertices $v, w \in V \setminus C$ such that $C$ is a $v$-$w$-vertex-cut in $H$. The* size *of a vertex-cut $C$ is its cardinality. A* minimum vertex-cut *is a vertex-cut of minimal size, and a* minimum $s$-$t$-vertex-cut *is an $s$-$t$-vertex-cut of minimal size. The* vertex-connectivity $\kappa(H)$ *of $H$ is the size of a minimum vertex-cut in $H$, and the $s$-$t$-*vertex-connectivity $\kappa_H(s,t)$ *of $H$ is the size of a minimum $s$-$t$-vertex-cut in $H$.*

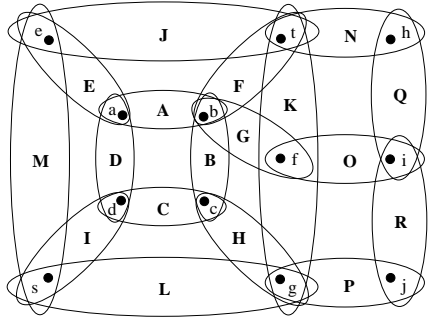 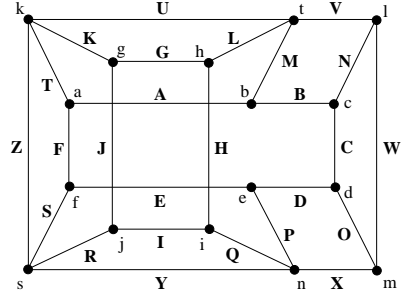

**Fig. 2.** Hypergraph example    **Fig. 3.** Graph example

*Example 2.* A minimum $s$-$t$-vertex-cut in the hypergraph in Fig. 2 is given by $\{d, e, g\}$, which separates vertex $s$ from vertex $t$. Similarly, a minimum $s$-$t$-vertex-cut in the graph in Fig. 3 is given by $\{f, j, k, n\}$, which also separates vertex $s$ from vertex $t$.

A minimum vertex-cut in the hypergraph in Fig. 2 is given by $\{g, i\}$, which separates vertex $j$ from all other vertices (except $g$ and $i$) in the hypergraph. Similarly, a minimum vertex-cut in the graph in Fig. 3 is given by $\{d, l, n\}$, which separates vertex $m$ from all other vertices (except $d$, $l$, and $n$) in the graph.

*Remark 2.* Consider hypergraph $H$ in Fig. 2 and graph $G$ in Fig. 3. From our examples above, we know that $\lambda(H) = \kappa(H) = 2$ and $\lambda(G) = \kappa(G) = 3$. Although edge-connectivity and vertex-connectivity are the same in both cases, it is easy to find hypergraphs where they do not coincide. However, for ordinary graphs, there is a well-known relationship between these invariants [10]: $\kappa(G) \leq \lambda(G) \leq \delta(G)$ for all graphs $G$, where $\delta(G)$ is the minimal degree over all vertices in $G$. Note that this result does not hold for hypergraphs in general. For example, consider a hypergraph $H$ consisting of two edges $e_1 = \{a, b, c\}$ and $e_2 = \{b, c, d\}$. Then we have $\lambda(H) = \delta(H) = 1 < 2 = \kappa(H)$.

## 4 The New Edge-Induced Vertex-Cut

In this section, we introduce our new edge-induced vertex-cuts and demonstrate their difference to edge-cuts and vertex-cuts described in the previous section. Intuitively, an edge-induced vertex-cut is a combination of an edge-cut and a vertex-cut in the sense that vertices are used to disconnect the hypergraph but the edges containing these vertices are used to measure the size of the cut. Recall the introduction for our motivation of investigating such cuts.

| Minimum Cuts | Hypergraph (Fig. 2) | Graph (Fig. 3) |
|---|---|---|
| *s-t-Edge-Cut* | $|\{I, L, M\}| = 3$ | $|\{R, S, Y, Z\}| = 4$ |
| *Edge-Cut* | $|\{P, R\}| = 2$ | $|\{O, W, X\}| = 3$ |
| *s-t-Vertex-Cut* | $|\{d, e, g\}| = 3$ | $|\{f, j, k, n\}| = 4$ |
| *Vertex-Cut* | $|\{g, i\}| = 2$ | $|\{d, l, n\}| = 3$ |
| *s-t-Edge-Induced Vertex-Cut* | $|\{E, H\}| = 2$ | $|\{H, P, T\}| = 3$ |
| *Edge-Induced Vertex-Cut* | $|\{K\}| = 1$ | $|\{M, P\}| = 2$ |

**Table 1.** Comparison of minimum cuts

**Definition 3 (Edge-Induced Vertex-Cut).** *Let $H = (V, E)$ be a hypergraph and $s, t \in V$ with $s \neq t$. A set $C \subseteq E$ is an s-t-edge-induced vertex-cut in $H$ if there exists $C' \subseteq \bigcup C$ such that $C'$ is an s-t-vertex-cut in $H$. A set $C \subseteq E$ is an* edge-induced vertex-cut *in $H$ if there are two distinct vertices $v, w \in V$ such that $C$ is a v-w-edge-induced vertex-cut in $H$. The* size *of an edge-induced vertex-cut $C$ is its cardinality. A* minimum edge-induced vertex-cut *is an edge-induced vertex-cut of minimal size, and a* minimum s-t-edge-induced vertex-cut *is an s-t-edge-induced vertex-cut of minimal size. The* edge-induced vertex-connectivity $\theta(H)$ *of $H$ is the size of a minimum edge-induced vertex-cut in $H$, and the s-t-edge-induced vertex-connectivity $\theta_H(s, t)$ of $H$ is the size of a minimum s-t-edge-induced vertex-cut in $H$.*

*Example 3.* A minimum $s$-$t$-edge-induced vertex-cut in the hypergraph in Fig. 2 is given by $\{E, H\}$, which separates vertex $s$ from vertex $t$ since all paths connecting $s$ and $t$ are going through vertices in $E \cup H$. Similarly, a minimum $s$-$t$-edge-induced vertex-cut in the graph in Fig. 3 is given by $\{H, P, T\}$.

A minimum edge-induced vertex-cut in the hypergraph in Fig. 2 is given by $\{K\}$, which separates the vertices $h$, $i$, and $j$ from the vertices $a$, $b$, $c$, $d$, $e$, and $s$ since all paths connecting these vertices are going through vertices in $K$. Similarly, a minimum edge-induced vertex-cut in the graph in Fig. 3 is given by $\{M, P\}$.

A comparison of all presented variants of minimum cuts concerning the hypergraph in Fig. 2 and the graph in Fig. 3 is now shown in Table 1. *Note that a minimum edge-induced vertex-cut is* not *just a cover of a minimum vertex-cut.* Moreover, note that our new edge-induced vertex-cut is in both cases smaller than the edge-cut and the vertex-cut. In the remainder of this section, we will show that this relationship holds in general. To this aim, let us first prove the following auxiliary result about the relationship between edge-cuts and edge-induced vertex-cuts.

**Lemma 1.** *Let $H = (V, E)$ be a hypergraph and $s, t \in V$. If $C \subseteq E$ is an s-t-edge-cut in $H$ such that none of the edges in $C$ contains both $s$ and $t$, then $C$ is an s-t-edge-induced vertex-cut in $H$.*

*Proof.* Let $C \subseteq E$ be an $s$-$t$-edge-cut in $H$, i.e., every path connecting $s$ and $t$ goes through edges in $C$, and assume that $\{s,t\} \not\subseteq e$ for all $e \in C$. Moreover, let $C' = \bigcup C \setminus \{s,t\}$. Now, w.l.o.g., consider any path $s = v_1, v_2, \ldots, v_k = t$ connecting $s$ and $t$. Then there exists a vertex $v_i$ with $1 \leq i < k$ and an edge $e \in C$ such that $v_i, v_{i+1} \in e$. Thus, since $\{s,t\} \not\subseteq e$ for all $e \in C$, we know that $v_i \in C'$ or $v_{i+1} \in C'$. Hence, every path connecting $s$ and $t$ goes through vertices in $C'$, i.e., $C' \subseteq \bigcup C$ is an $s$-$t$-vertex-cut in $H$. Consequently, $C$ is an $s$-$t$-edge-induced vertex-cut in $H$. □

We are now able to prove our first two main results. Let us start with the relationship between the $s$-$t$-edge-connectivity, the $s$-$t$-vertex-connectivity, and the $s$-$t$-edge-induced vertex-connectivity of hypergraphs. To this aim, recall Remark 2.

**Theorem 1.** *Let $H = (V, E)$ be a hypergraph and $s, t \in V$ such that $\theta_H(s,t)$ is defined. Then it follows that $\theta_H(s,t) \leq \min(\kappa_H(s,t), \lambda_H(s,t))$.*

*Proof.* To show that $\theta_H(s,t) \leq \min(\kappa_H(s,t), \lambda_H(s,t))$, we show first that $\theta_H(s,t) \leq \kappa_H(s,t)$. To this aim, let $C \subseteq V$ be an $s$-$t$-vertex-cut of size $\kappa_H(s,t)$ in $H$. Now, we construct $C' \subseteq E$ in the following way: Starting at the empty set, we add for each vertex $v \in C$ an edge $e \in E$ with $v \in e$ to $C'$. Thus, $C \subseteq \bigcup C'$ and $|C'| \leq \kappa_H(s,t)$. Hence, $C'$ is an $s$-$t$-edge-induced vertex-cut in $H$, and it holds that $\theta_H(s,t) \leq |C'| \leq \kappa_H(s,t)$.

Now, let us show that $\theta_H(s,t) \leq \lambda_H(s,t)$. To this aim, let $C \subseteq E$ be an $s$-$t$-edge-cut of size $\lambda_H(s,t)$ in $H$. Note that there cannot be an edge $e \in C$ such that $\{s,t\} \subseteq e$; otherwise, $\theta_H(s,t)$ would be undefined. Hence, by Lemma 1, we know that $C$ is an $s$-$t$-edge-induced vertex-cut in $H$, and it holds that $\theta_H(s,t) \leq |C| = \lambda_H(s,t)$. □

Now, let us prove an analogous result for the unrestricted case, i.e., the relationship between the edge-connectivity, the vertex-connectivity, and the edge-induced vertex-connectivity. Note that Theorem 2 does not follow from Theorem 1, since $\theta(H)$ being defined does not imply that $\theta_H(s,t)$ for some fixed vertices $s$ and $t$ is defined.

**Theorem 2.** *Let $H = (V, E)$ be a hypergraph such that $\theta(H)$ is defined. Then it follows that $\theta(H) \leq \min(\kappa(H), \lambda(H))$.*

*Proof.* To show that $\theta(H) \leq \min(\kappa(H), \lambda(H))$, we show first that $\theta(H) \leq \kappa(H)$. To this aim, let $C \subseteq V$ be a vertex-cut of size $\kappa(H)$ in $H$. Now, we construct $C' \subseteq E$ in the following way: Starting at the empty set, we add for each vertex $v \in C$ an edge $e \in E$ with $v \in e$ to $C'$. Thus, $C \subseteq \bigcup C'$ and $|C'| \leq \kappa(H)$. Hence, $C'$ is an edge-induced vertex-cut in $H$, and it holds that $\theta(H) \leq |C'| \leq \kappa(H)$.

Now, let us show that $\theta(H) \leq \lambda(H)$. To this aim, let $C \subseteq E$ be an edge-cut of size $\lambda(H)$ in $H$. We have now to distinguish between two cases:

(i) Let us first assume that for all edge-cuts $C'$ of size $\lambda(H)$ and all vertices $v, w \in V$ belonging to different components induced by $C'$, there exists an edge $e \in C'$ with $v, w \in e$. Intuitively, this means that there is no pair $v$ and $w$ disconnected by some edge-cut of size $\lambda(H)$ such that $v$ and $w$ can be disconnected by a vertex-cut (since there is always an edge connecting $v$ and $w$). Thus, we know that a vertex-cut must disconnect two vertices $v$ and $w$ belonging to the same component w.r.t. $C$. Note that such a vertex-cut must exist; otherwise, $\theta(H)$ would be undefined. Let $C' \subseteq V$ be such a vertex-cut, i.e., every path connecting $v$ and $w$ goes through vertices in $C'$. By assumption, however, we know that for each $x \in C'$ there exists $e \in C$ such that $x \in e$, that is $C' \subseteq \bigcup C$. Hence, $C$ is an edge-induced vertex-cut in $H$, and it holds that $\theta(H) \leq |C| = \lambda(H)$.

(ii) Otherwise, we can assume w.l.o.g. that there are two vertices $v, w \in V$ belonging to different components induced by $C$ such that $\{v, w\} \not\subseteq e$ for all $e \in C$. Hence, by Lemma 1, we know that $C$ is an edge-induced vertex-cut in $H$, and it holds that $\theta(H) \leq |C| = \lambda(H)$. □

From the relationship between edge-connectivity and vertex-connectivity of graphs (recall Remark 2) and Theorem 2, it follows immediately that for ordinary graphs $G$ it holds that $\theta(G) \leq \kappa(G) \leq \lambda(G) \leq \delta(G)$. Similarly, it holds that $\theta_G(s,t) \leq \kappa_G(s,t) \leq \lambda_G(s,t) \leq \min(\delta_G(s), \delta_G(t))$.

## 5 The Complexity of Computing Minimum Cuts

In this section, we first give a short overview of the time complexity of known algorithms for computing minimum edge-cuts and minimum vertex-cuts. Afterwards, we prove the complexity of deciding whether the size of a minimum edge-induced vertex-cut is less than or equal to a given integer $k$. In both parts we distinguish between hypergraphs in general and their restriction to ordinary graphs. For notational convenience, let $n$ denote the number of vertices and $m$ denote the number of edges in a given hypergraph $H = (V, E)$.

Several polynomial-time algorithms for computing minimum $s$-$t$-edge-cuts in graphs have been presented in the literature; most of them are based on network flow techniques. One such well-known algorithm is due to Goldberg and Tarjan [6] and runs in time $\mathcal{O}(n^2 \sqrt{m})$. We will now shortly describe how to reduce in polynomial time all remaining edge-cuts and vertex-cuts of hypergraphs and graphs to $s$-$t$-edge-cuts of graphs, which results in polynomial-time algorithms for all these cut problems. Note that this does not contradict the results of Ihler et al. [9], who showed that computing minimum edge-cuts in hypergraphs cannot be reduced to computing minimum edge-cuts in (undirected) graphs without using negative weights; however, in our reduction, we use *directed* graphs. Moreover, note that there exist algorithms in the literature with much better time complexity. In particular, our runtime bounds hold for the case of weighted hypergraphs; in the case of unweighted hypergraphs, there exist faster algorithms.

First note that computing a minimum edge-cut can always be reduced to $n-1$ computations of a minimum $s$-$t$-edge-cut. Thus, computing minimum edge-cuts in graphs can be done in time $\mathcal{O}(n^3 \sqrt{m})$. Moreover, by transforming vertices into edges [4], we are able to reduce the minimum $s$-$t$-vertex-cut problem in a graph with $n$ vertices and $m$ edges to a minimum $s$-$t$-edge-cut problem in a graph with $2n$ vertices and $n + 2m$ edges. Thus, we obtain an algorithm for computing minimum $s$-$t$-vertex-cuts in graphs in time $\mathcal{O}(n^2 \sqrt{n+m})$. Similarly, by reducing the computation of minimum vertex-cuts in graphs to the computation of minimum edge-cuts in graphs, we obtain for the minimum vertex-cut problem of graphs a time complexity of $\mathcal{O}(n^3 \sqrt{n+m})$.

So we have seen that all minimum edge-cuts and vertex-cuts for graphs can be computed in polynomial time. In the case of hypergraphs, our encoding becomes more expensive but remains polynomial: Let $p$ denote the sum over all cardinalities of edges in $E$, i.e., $p = \sum_{e \in E} |e|$. Now, we transform each hyperedge into a star, i.e., for each hyperedge we introduce a new vertex and connect this vertex with all vertices in the corresponding hyperedge [8]. Afterwards, the vertices corresponding to a hyperedge in the hypergraph or the vertices corresponding to vertices in the hypergraph (depending on whether we want to compute edge-cuts or vertex-cuts) can be transformed into edges [4], which allows us to apply our algorithms for computing minimum edge-cuts in graphs. In particular, computing a minimum edge-cut in a hypergraph can be reduced to computing a minimum edge-cut in a graph with $n + 2m$ vertices and $m + 2p$ edges, and computing a minimum vertex-cut in a hypergraph can be reduced to computing a minimum edge-cut in a graph with $2n + m$ vertices and $n + 2p$ edges. In this way, we obtain a runtime of $\mathcal{O}((n+m)^2 \sqrt{p})$ for computing minimum $s$-$t$-edge-cuts and minimum $s$-$t$-vertex-cuts in hypergraphs, and a runtime of $\mathcal{O}((n+m)^3 \sqrt{p})$ for computing minimum edge-cuts and minimum vertex-cuts in hypergraphs.

| Minimum Cuts | Hypergraphs | Graphs |
|---|:---:|:---:|
| *s-t-Edge-Cut* | in P | in P |
| *Edge-Cut* | in P | in P |
| *s-t-Vertex-Cut* | in P | in P |
| *Vertex-Cut* | in P | in P |
| *s-t-Edge-Induced Vertex-Cut* | NP-complete, W[2]-hard | NP-complete |
| *Edge-Induced Vertex-Cut* | NP-complete, W[2]-hard | NP-complete |

**Table 2.** Complexity overview

Let us finally mention that there exists also a very simple and efficient algorithm for computing minimum edge-cuts which is not based on network flow techniques. This algorithm is due to Stoer and Wagner [15] and has a runtime of $\mathcal{O}(nm + n^2 \log n)$. Its generalization to hypergraphs [11,12] has a runtime of $\mathcal{O}(np + n^2 \log n)$.

The interested reader is referred to other sources [1,10] for further mincut algorithms and their complexity. Table 2 gives an overview of the time complexity of the above considered minimum edge-cut and vertex-cut problems for weighted hypergraphs and graphs. Moreover, it shows that our new edge-induced vertex-cut decision problem is NP-complete even for ordinary graphs, which we will now prove. Let us start with the restricted variant of $s$-$t$-edge-induced vertex cuts.

**Theorem 3.** *Given a graph $G = (V, E)$, two distinct vertices $s, t \in V$, and a positive integer $k$. Deciding whether $G$ has an $s$-$t$-edge-induced vertex-cut of size at most $k$ is* NP-*complete.*

*Proof.* Clearly the problem is in NP since it can be checked in polynomial time whether a guessed cut has size at most $k$ and whether it disconnects $s$ and $t$. To show that the problem is NP-hard, we give a reduction from 3SAT.

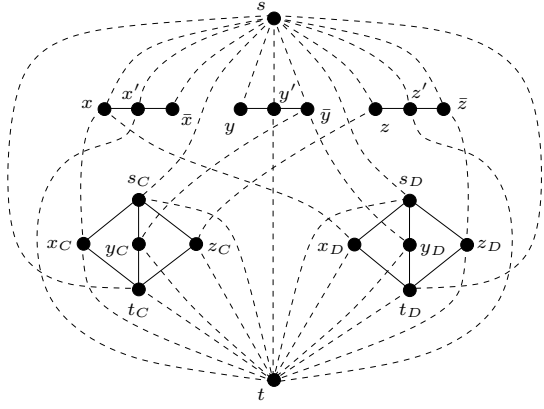

**Fig. 4.** Encoding of the 3SAT instance $\{\{x, \neg y, z\}, \{x, \neg y, \neg z\}\}$

Consider an instance of 3SAT given by a set $F$ of clauses; each clause $C \in F$ is a set containing exactly three literals, that is, negated or unnegated propositional variables. We write $var(C)$ to denote the set of variables occurring in clause $C$, and we set $var(F) = \bigcup_{C \in F} var(C)$. Now, let us construct a graph $G = (V, E)$ from this instance as exemplified in Fig. 4. The set $V$ of vertices is given by $V_v \cup V_c \cup \{s, t\}$ with $s, t \notin V_v \cup V_c$, where $V_v = \{x, x', \bar{x} \mid x \in var(F)\}$ and $V_c = \{s_C, t_C, x_C \mid C \in F, x \in var(C)\}$. The set $E$ of edges is given by $E_0 \cup E_1$, where $E_0 = \{\{x, x'\}, \{x', \bar{x}\} \mid x \in var(F)\} \cup \{\{s_C, x_C\}, \{x_C, t_C\} \mid C \in F, x \in var(C)\}$ and $E_1 = \{\{s, x\}, \{s, x'\}, \{s, \bar{x}\}, \{x', t\} \mid x \in var(F)\} \cup \{\{s, s_C\}, \{s_C, t\}, \{s, t_C\}, \{t_C, t\} \mid C \in F\} \cup \{\{x_C, t\} \mid C \in F, x \in var(C)\} \cup \{\{x, x_C\} \mid C \in F, x \in C\} \cup \{\{\bar{x}, x_C\} \mid C \in F, \neg x \in C\}$. In Fig. 4, the edges in $E_0$ are indicated by solid lines and the edges in $E_1$ are indicated by dashed lines. We will now show that $F$ is satisfiable if and only if there exists an $s$-$t$-edge-induced vertex-cut of size at most $k = |var(F)| + 2|F|$ in $G$.

For the *only if* direction assume that there exists a satisfying assignment $\alpha$ of $F$. We will now construct an $s$-$t$-edge-induced vertex-cut of size $k$ from this assignment in the following way: (i) For each atom $x \in var(F)$ we select the edge $\{x, x'\}$ if $x$ is *true* under $\alpha$, and we select the edge $\{x', \bar{x}\}$ if $x$ is *false* under $\alpha$. (ii) For each clause $C \in F$ let $l \in C$ be a literal that evaluates to *true* (note that there must be at least one such literal) and let $x, y \in var(C)$ be the variables of the other two literals in $C$. Then we select the edges $\{s_C, x_C\}$ and $\{y_C, t_C\}$. Now, it is easy to verify that every path from $s$ to $t$ goes through vertices in the selected edges. Hence, they represent an $s$-$t$-edge-induced vertex-cut of size $k$.

For the *if* direction consider an $s$-$t$-edge-induced vertex-cut in $G$ of size at most $k$. Note that we can assume w.l.o.g. that this cut consists of edges in $E_0$ only. To see this, note that every edge incident with $s$ or $t$ can be replaced by an arbitrary adjacent edge in $E_0$ because $s$ and $t$ are by definition not allowed to disconnect the graph. In particular, for each $x \in var(F)$, there must be an edge incident with $x'$ in the cut; otherwise the path $s, x', t$ connects $s$ and $t$. Since each edge in $E_1$ incident with $x'$ is also incident with $s$ or $t$, it can be replaced by $\{x, x'\}$ or $\{x', \bar{x}\}$, both in $E_0$. Moreover, for each $C \in F$ with $var(C) = \{x, y, z\}$, there must be an edge incident with $s_C$ and an edge incident with $t_C$ in the cut; otherwise the path $s, s_C, t$ resp. $s, t_C, t$ connects $s$ and $t$. Since each edge in $E_1$ incident with $s_C$ resp. $t_C$ is also incident with $s$ or $t$, it can be replaced by $\{s_C, x_C\}, \{s_C, y_C\}$, or $\{s_C, z_C\}$ resp. $\{x_C, t_C\}, \{y_C, t_C\}$, or $\{z_C, t_C\}$, all of them in $E_0$. By our upper bound $k$, however, we know that no other edges are in the cut. In particular, this means that for each vertex $x \in var(F)$ either $\{x, x'\}$ or $\{x', \bar{x}\}$ must be in the cut but not both of them. Now, we obtain a satisfying assignment $\alpha$ of $F$ by assigning *true* to $x$ if $\{x, x'\}$ is in the cut, and by assigning *false* to $x$ if $\{x', \bar{x}\}$ is in the cut. To see that $\alpha$ is indeed a satisfying assignment, consider any clause $C \in F$. By our upper bound $k$, we know that there exists $x \in var(C)$ such that neither $\{s_C, x_C\}$ nor $\{x_C, t_C\}$ is in the cut. This, however, implies that (i) $\{x, x'\}$ is in the cut if $x \in C$ or (ii) $\{x', \bar{x}\}$ is in the cut if $\neg x \in C$; otherwise the path $s, x, x_C, t$ resp. $s, \bar{x}, x_C, t$ connects $s$ and $t$. Since $x$ is *true* under $\alpha$ in case (i) and *false* under $\alpha$ in case (ii), we know that $C$ evaluates to *true*. Hence, $\alpha$ satisfies $F$ and so $F$ is satisfiable. □

Now, let us consider the case of unrestricted edge-induced vertex-cuts. The proof of the following theorem is a generalization of the proof of Theorem 3.

**Theorem 4.** *Given a graph $G$ and a positive integer $k$. Deciding whether $G$ has an edge-induced vertex-cut of size at most $k$ is* NP-*complete.*

*Proof.* Clearly the problem is in NP since it can be checked in polynomial time whether a guessed cut has size at most $k$ and whether it disconnects the graph. To show that the problem is NP-hard, we give a reduction from 3SAT.

Consider an instance of 3SAT given by a set $F$ of clauses; each clause $C \in F$ is a set containing exactly three literals, that is, negated or unnegated propositional variables. We write $var(C)$ to denote the set of variables occurring in clause $C$, and we set $var(F) = \bigcup_{C \in F} var(C)$. Now, let us construct a graph $G = (V, E)$ from this instance as exemplified in Fig. 4. The set $V$ of vertices is given by $V_v \cup V_c \cup S \cup T$, where $V_v$ and $V_c$ are the same as in Theorem 3, and $S$ and $T$ are sets of new vertices such that $|S| = |T| = |var(F)| + 2|F| + 1$, $S \cap T = \emptyset$, and $(S \cup T) \cap (V_v \cup V_c) = \emptyset$. The set $E$ of edges is given by $E_0 \cup E_1$, where $E_0$ is the same as in Theorem 3 and $E_1 = \{\{u, x\}, \{u, x'\}, \{u, \bar{x}\}, \{x', v\} \mid u \in S, v \in T, x \in var(F)\} \cup \{\{u, s_C\}, \{s_C, v\}, \{u, t_C\}, \{t_C, v\} \mid u \in S, v \in T, C \in F\} \cup \{\{x_C, v\} \mid C \in F, x \in var(C), v \in T\} \cup \{\{x, x_C\} \mid C \in F, x \in C\} \cup \{\{\bar{x}, x_C\} \mid C \in F, \neg x \in C\}$. In Fig. 4, we assume that $S = \{s\}$ and $T = \{t\}$; the edges in $E_0$ are indicated by solid lines and the edges in $E_1$ are indicated by dashed lines. We will now show that $F$ is satisfiable if and only if there exists an edge-induced vertex-cut of size at most $k = |var(F)| + 2|F|$ in $G$.

For the *only if* direction assume that there exists a satisfying assignment $\alpha$ of $F$. Moreover, let $s \in S$ and $t \in T$. Now we construct an $s$-$t$-edge-induced vertex-cut of size $k$ from $\alpha$ in the same way as in Theorem 3. Since every $s$-$t$-edge-induced vertex-cut is also an edge-induced vertex-cut, we are done.

For the *if* direction consider a $v$-$w$-edge-induced vertex-cut in $G$ of size at most $k$. Since $|S| = k + 1$ and $|T| = k + 1$, we know that there is at least one $s \in S$ and one $t \in T$ such that neither $s$ nor $t$ is incident with an edge in the cut. Now, assume for

the sake of contradiction that there is a path $s = v_1, v_2, \ldots, v_k = t$ not going through vertices in the cut. Thus, $v_2$ and $v_{k-1}$ are not incident with any edge in the cut. Since all vertices in $S$ are adjacent to $v_2$ and all vertices in $T$ are adjacent to $v_{k-1}$ by construction, we know that there is a path from $s$ to all vertices in $S$ and from $t$ to all vertices in $T$. Moreover, since $s$ is adjacent to all vertices in $V_v$ and $t$ is adjacent to all vertices in $V_c$ by construction, we know that all pairs of vertices in $S \cup V_v$ and all pairs of vertices in $T \cup V_c$ are connected. Consequently, since there is a path from $s$ to $t$, we know that all pairs of vertices in $V$ are connected. This, however, contradicts our assumption that $v$ and $w$ are disconnected by the cut. Hence, $s$ and $t$ must be disconnected as well, i.e., our $v$-$w$-edge-induced vertex-cut is also an $s$-$t$-edge-induced vertex-cut. Now, we can apply similar arguments as in Theorem 3 to show that $F$ is satisfiable. □

Note that in the more general setting of *hypergraphs*, NP-hardness of the $s$-$t$-edge-induced vertex-cut problem follows trivially from Theorem 3 and NP-hardness of the edge-induced vertex-cut problem follows trivially from Theorem 4. Moreover, both problems are clearly in NP since it can be checked in polynomial time whether a guessed cut has size at most $k$ and whether it disconnects the given hypergraph. Hence, we immediately obtain NP-completeness in the case of hypergraphs as well.

### 5.1 Parameterized Complexity of Edge-Induced Vertex-Cuts

The framework of *parameterized complexity* provides an adequate concept and tools for studying the question whether a parameter $k$ of a decision problem allows algorithms with time complexity

$$\text{(i)} \ \mathcal{O}(\|I\|^{\mathcal{O}(f(k))}) \qquad \text{or} \qquad \text{(ii)} \ \mathcal{O}(f(k)\|I\|^{\mathcal{O}(1)}),$$

where $\|I\|$ denotes the input size of the problem instance $I$ and $f$ denotes a computable function. The runtime of type (i) is polynomial when $k$ is considered as a constant. However, since $k$ appears in the exponent, such algorithms become impractical—even if $k$ is small—when large instances are considered. In contrast, the runtime of type (ii) is significantly better since the polynomial does not depend on $k$, and so considering larger and larger classes w.r.t. $k$ does not increase the order of the polynomial.

Parameterized complexity was initiated by Downey and Fellows in the late 1980s and has become an important branch of algorithm design and analysis [3,13,5]. It turned out that the distinction between tractability of type (i) and type (ii) is a robust indication of problem hardness. XP denotes the class of all parameterized decision problems which can be solved in runtime of type (i). A *fixed-parameter algorithm* is an algorithm that achieves a runtime of type (ii). A parameterized problem is *fixed-parameter tractable* if it can be solved by a fixed-parameter algorithm. FPT denotes the class of all fixed-parameter tractable decision problems.

Parameterized complexity offers a *completeness theory*, similar to the theory of NP-completeness, that allows the accumulation of strong theoretical evidence that a parameterized problem is *not* fixed-parameter tractable. This completeness theory is based on the *weft hierarchy* of complexity classes $W[1], W[2], \ldots, W[P]$. Each class is the equivalence class of certain parameterized satisfiability problems under *fpt-reductions*.

Let $\Pi$ and $\Pi'$ be two parameterized problems. An *fpt-reduction R from $\Pi$ to $\Pi'$* is a many-to-one transformation from $\Pi$ to $\Pi'$, such that (i) $(I, k) \in \Pi$ if and only if $(I', k') \in \Pi'$ with $k' \leq g(k)$ for a fixed computable function $g$ and (ii) $R$ is of complexity $\mathcal{O}(f(k) \|I\|^{\mathcal{O}(1)})$ for a computable function $f$. The above classes form the chain

$$\text{FPT} \subseteq \text{W}[1] \subseteq \text{W}[2] \subseteq \cdots \subseteq \text{W}[P] \subseteq \text{XP}$$

where all inclusions are assumed to be proper. A parameterized analog of Cook's Theorem gives strong evidence to assume that $\text{FPT} \neq \text{W}[1]$; it is known that $\text{FPT} \neq \text{XP}$ [3]. Although XP contains problems which are very unlikely to be fixed-parameter tractable, it is often a significant improvement to show that a problem belongs to this class, in contrast to, e.g., $k$-SAT which is NP-complete for every constant $k \geq 3$.

The following parameterized set cover-problem is W[2]-complete [3]; this problem is the basis for the hardness results considered in the sequel.

**SET COVER**
*Instance:* A finite family of finite sets $S$ and a positive integer $k$.
*Parameter:* $k$.
*Question:* Is there a subset $R \subseteq S$ with $|R| \leq k$ whose union is all elements in the union of $S$?

**Theorem 5.** *Given a hypergraph $H = (V, E)$, two distinct vertices $s, t \in V$, and a positive integer $k$. Deciding whether $H$ has an $s$-$t$-edge-induced vertex-cut of size at most $k$ is* W[2]*-hard w.r.t. parameter $k$.*

*Proof.* Consider an instance of SET COVER given by a finite family of finite sets $S$ and a positive integer $k$. Now, let us construct a hypergraph $H = (V, E)$ from this instance in the following way: The set $V$ of vertices is given by the set $\bigcup S$ together with two new vertices $s$ and $t$ not in $\bigcup S$, i.e., $V = \bigcup S \cup \{s, t\}$ with $s, t \notin \bigcup S$. The set $E$ of hyperedges is given by the set $S$ together with $2 |\bigcup S|$ new hyperedges connecting $s$ and $t$ with each vertex in $\bigcup S$, i.e., $E = S \cup \{\{s, x\}, \{x, t\} \mid x \in \bigcup S\}$.

We will now show that there exists $R \subseteq S$ with $|R| \leq k$ such that $\bigcup R = \bigcup S$ if and only if there exists an $s$-$t$-edge-induced vertex-cut of size at most $k$ in $H$.

For the *only if* direction let $R \subseteq S$ with $|R| \leq k$ such that $\bigcup R = \bigcup S$. Note that by construction every path connecting $s$ and $t$ goes through vertices in $\bigcup S$. Hence, since $\bigcup R = \bigcup S$, we know that $R$ is an $s$-$t$-edge-induced vertex-cut of size at most $k$ in $H$.

For the *if* direction let $C \subseteq E$ be an $s$-$t$-edge-induced vertex-cut of size at most $k$ in $H$, i.e., every path connecting $s$ and $t$ goes through vertices in $\bigcup C$. Now, assume for the sake of contradiction that there exists $x \in \bigcup S \setminus \bigcup C$. This, however, implies that there exists a path $s, x, t$ connecting $s$ and $t$, which contradicts that $C$ is an $s$-$t$-edge-induced vertex-cut. Hence, $\bigcup S \subseteq \bigcup C$. Now, let us construct $R$ in the following way: Starting at the empty set, we add all edges in $C$ that contain neither $s$ nor $t$ to $R$, i.e., $R = C \cap S$. All remaining edges $e \in C \setminus R$ must then be of the form $\{s, x\}$ or $\{x, t\}$ with $x \in \bigcup S$. By construction, we know that for each such $x$ there exists an edge $e' \in S$ with $x \in e'$. Thus, for each $e \in C \setminus R$ we add an edge $e' \in S$ with $e \cap \bigcup S \subseteq e'$ to $R$. Hence, we have $R \subseteq S$, $|R| \leq |C| \leq k$, and $\bigcup R = \bigcup C \cap \bigcup S = \bigcup S$. □

Now, let us consider the case of unrestricted edge-induced vertex-cuts. The proof of the following theorem is a generalization of the proof of Theorem 5.

**Theorem 6.** *Given a hypergraph $H$ and a positive integer $k$. Deciding whether $H$ has an edge-induced vertex-cut of size at most $k$ is* W[2]*-hard w.r.t. parameter $k$.*

*Proof.* Consider an instance of SET COVER given by a finite family of finite sets $S$ and a positive integer $k$. Now, let us construct a hypergraph $H = (V, E)$ from this instance in the following way: The set $V$ of vertices is given by the set $\bigcup S$ together with $|S|+1$ new vertices $v_0, v_1, v_2, \ldots, v_{|S|}$ not in $\bigcup S$, i.e., $V = \bigcup S \cup \{v_i \mid 0 \leq i \leq |S|\}$ with $v_i \notin \bigcup S$ for all $0 \leq i \leq |S|$. The set $E$ of hyperedges is given by the set $S$ together with $(|S| + 1) \cdot |\bigcup S|$ new hyperedges connecting $v_0, v_1, v_2, \ldots, v_{|S|}$ with each vertex in $\bigcup S$, i.e., $E = S \cup \{\{x,y\} \mid x \in V \setminus \bigcup S, \ y \in \bigcup S\}$.

We will now show that there exists $R \subseteq S$ with $|R| \leq k$ such that $\bigcup R = \bigcup S$ if and only if there exists an edge-induced vertex-cut of size at most $k$ in $H$.

For the *only if* direction let $R \subseteq S$ with $|R| \leq k$ such that $\bigcup R = \bigcup S$. Moreover, let $s, t \in V \setminus \bigcup S$ with $s \neq t$. Now we can apply the same argument as in Theorem 5. Hence, $R$ is an edge-induced vertex-cut of size at most $k$ in $H$.

For the *if* direction let $C \subseteq E$ be an edge-induced vertex-cut of size at most $k$ in $H$, i.e., there exist distinct vertices $s, t \in V$ such that every path connecting $s$ and $t$ goes through vertices in $\bigcup C$. Now, let us distinguish between two cases: If $|S| \leq |C| \leq k$, we choose $R = S$ and are done. Otherwise, if $|S| > |C|$, we know that there exists $x \in V \setminus \bigcup S$ such that $x \notin \bigcup C$, since by construction it holds that $|V \setminus \bigcup S| > |S| > |C|$ and there is no edge in $E$ containing more than one vertex in $V \setminus \bigcup S$. This, however, implies that $s, t \in V \setminus \bigcup S$. Otherwise, there would be two possibilities: First, $s \in \bigcup S$ and $t \in \bigcup S$, which implies $\{s, x\} \in E$ and $\{x, t\} \in E$, i.e., there exists a path $s, x, t$. Second, $s \in V \setminus \bigcup S$ and $t \in \bigcup S$ (resp. $s \in \bigcup S$ and $t \in V \setminus \bigcup S$), which implies $\{s, t\} \in E$, i.e., there exists a path $s, t$. In both cases, we have a path connecting $s$ and $t$ which does not go through vertices in $\bigcup C$. This, however, contradicts that $C$ is an $s$-$t$-edge-induced vertex-cut. Based on the fact $s, t \in V \setminus \bigcup S$, we can now apply similar arguments as in Theorem 5 to construct an appropriate set $R$. □

Note that Theorems 5 and 6 prove also NP-hardness of computing minimum edge-induced vertex-cuts in hypergraphs. However, these proofs do not work for ordinary graphs, since SET COVER is polynomially solvable via matching algorithms [14] if the size of the sets is bounded by 2. On the other hand, the reductions used for showing Theorems 3 and 4 are not fpt-reductions and do not establish W[2]-hardness.

## 6   Applications of Edge-Induced Vertex-Cuts

Next we describe applications of edge-induced vertex-cuts, as mentioned in the introduction, in more detail. Due to the negative results concerning the time complexity of computing minimum edge-induced vertex-cuts in Section 5, we believe that important applications will motivate further research for approximating edge-induced vertex-cuts efficiently.

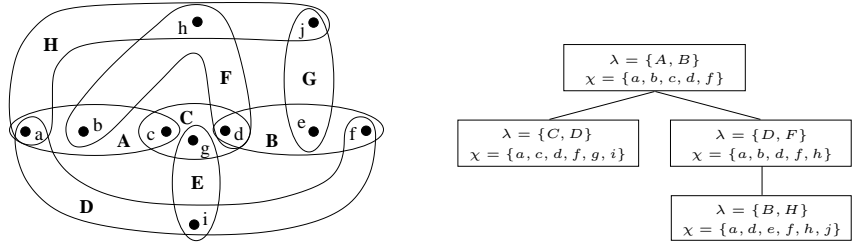

**Fig. 5.** Example of a hypergraph and its generalized hypertree decomposition

Our first application concerns network reliability, i.e., the number of network components that have to fail in order to disconnect a network. For example, assume the hypergraph in Fig. 2 models a telecommunication network and we are interested in how resistent our network is against attacks. In particular, let us assume that vertices $s$ and $t$ represent headquarters and we want to know the critical network components which allow an attacker to disconnect $s$ and $t$ with a minimum of effort. From Table 1 we know that an attacker has to cut at least three communication lines (e.g., $I$, $L$, and $M$) or to destroy at least three communication stations (e.g., $d$, $e$, and $g$) in order to disconnect $s$ and $t$. However, by our minimum $s$-$t$-edge-induced vertex-cut we know that it suffices to apply high voltage on the communication lines $E$ and $H$ in order to destroy the directly connected stations $a$, $c$, $e$, and $g$ which thereby disconnect $s$ and $t$. Thus, an attacker needs access to only two network components, namely $E$ and $H$, to disconnect our headquarters. Security decisions may now be based on this information.

Our second application concerns hypertree decompositions [7]—our original motivation for investigating edge-induced vertex-cuts. A *hypertree* $(T, \chi, \lambda)$ for a hypergraph $H$ is a tree $T$ with two labeling functions $\chi$ and $\lambda$, where $\chi : V(T) \longrightarrow 2^{V(H)}$ associates to each tree node a set of vertices of $H$ and $\lambda : V(T) \longrightarrow 2^{E(H)}$ associates to each tree node a set of edges of $H$. For each subtree $T'$ of $T$, we define $\chi(T') = \bigcup_{p \in V(T')} \chi(p)$, and for each $p \in V(T)$, we denote the subtree of $T$ rooted at $p$ by $T_p$.

A *generalized hypertree decomposition* of a hypergraph $H$ is a hypertree $T$ for $H$ satisfying three conditions: (i) $\forall e \in E(H)\, \exists p \in V(T) : e \subseteq \chi(p)$, (ii) ("Connectedness Condition") $\forall v \in V(H)$ : the set $\{p \in V(T)\,|\, v \in \chi(p)\}$ induces a (connected) subtree of $T$, and (iii) $\forall p \in V(T) : \chi(p) \subseteq \bigcup \lambda(p)$. A *hypertree decomposition* is a generalized hypertree decomposition satisfying the additional condition: (iv) $\forall p \in V(T) : \bigcup \lambda(p) \cap \chi(T_p) \subseteq \chi(p)$. The *width* of a (generalized) hypertree decomposition is given by $\max_{p \in V(T)} |\lambda(p)|$, and the *(generalized) hypertree-width* of a hypergraph is the minimal width over all its (generalized) hypertree decompositions.

Fig. 5 shows an example of a hypergraph and its generalized hypertree decomposition of width 2. This generalized hypertree decomposition is optimal, i.e., the generalized hypertree-width of the hypergraph is 2. Now, consider the $\lambda$-sets of the hypertree. It is easy to verify that they are all edge-induced vertex-cuts; in particular, they all disconnect vertices $i$ and $j$. In fact, they are *minimum* edge-induced vertex-cuts since there is no edge-induced vertex-cut of size 1 in our hypergraph. We will now show that the $\lambda$-sets of hypertree decompositions are always edge-induced vertex-cuts.

**Theorem 7.** *Let $(T, \chi, \lambda)$ be a generalized hypertree decomposition of a hypergraph $H$ and $p, q \in V(T)$ be two hypertree nodes such that $p$ is the parent of $q$. Then for all $s, t \in \chi(T) \setminus (\chi(p) \cap \chi(q))$ such that $s \notin \chi(T_q)$ and $t \in \chi(T_q)$, it follows that $\lambda(p)$ and $\lambda(q)$ are s-t-edge-induced vertex-cuts in $H$.*

*Proof.* For simplicity, let us define $X = \chi(T) \setminus \chi(T_q)$ and $Y = \chi(T_q)$. Since $t \in Y$ but $t \notin \chi(p) \cap \chi(q)$, we know by the Connectedness Condition that $t \notin X$. So we have $s \in X$ but $t \notin X$, and $t \in Y$ but $s \notin Y$. Now, consider any path $s = v_1, v_2, \ldots, v_k = t$ connecting $s$ and $t$ in $H$. Then, since $s \in X$ and $t \in Y$, there must exist a vertex $v_i$ with $1 \leq i < k$ and an edge $e \in E$ such that $v_i, v_{i+1} \in e$ and $v_i \in X$ and $v_{i+1} \in Y$. By the first condition of a generalized hypertree decomposition, this implies that there exists $r \in V(T)$ such that $e \subseteq \chi(r)$. Now, we have to distinguish between two cases: (i) If $r \in V(T_q)$, we know that $v_i \in Y$ and thus $v_i \in X \cap Y$. By the Connectedness Condition, however, this implies that $v_i \in \chi(p) \cap \chi(q)$. Moreover, since $v_i \in \chi(p) \cap \chi(q)$ and $s, t \notin \chi(p) \cap \chi(q)$, we know that $v_i \neq s$ and $v_i \neq t$. Thus, $\chi(p)$ and $\chi(q)$ are s-t-vertex-cuts in $H$. By the third condition of a generalized hypertree decomposition, we know that $\chi(p) \subseteq \bigcup \lambda(p)$ and $\chi(q) \subseteq \bigcup \lambda(q)$. Hence, $\lambda(p)$ and $\lambda(q)$ are s-t-edge-induced vertex-cuts in $H$. (ii) The case of $r \in V(T) \setminus V(T_q)$ is completely analogous. $\square$

To show an immediate consequence of this result, we need the following definition.

**Definition 4** ($\chi$-reduced). *A hypertree $(T, \chi, \lambda)$ is $\chi$-reduced if for all adjacent hypertree nodes $p, q \in V(T)$ it holds that $\chi(p) \not\subseteq \chi(q)$ and $\chi(q) \not\subseteq \chi(p)$.*

Every hypertree decomposition of width $k$ can be easily transformed into a $\chi$-reduced hypertree decomposition of width $k$: For all adjacent hypertree nodes $p$ and $q$ satisfying $\chi(p) \subseteq \chi(q)$, remove node $p$ from the hypertree and connect all its adjacent nodes $p'$ such that $p' \neq q$ with $q$. It is easy to verify that all hypertree conditions remain satisfied during such a transformation. For example, our hypertree decomposition in Fig. 5 is already $\chi$-reduced. Moreover, let us say a hypertree is *non-trivial* if it consists of at least two hypertree nodes.

It is easy to see that in every non-trivial $\chi$-reduced hypertree there exist vertices $s$ and $t$ satisfying the conditions of Theorem 7. Thus, we obtain

**Corollary 1.** *Let $(T, \chi, \lambda)$ be a non-trivial $\chi$-reduced hypertree decomposition of a hypergraph $H$. Then for every hypertree node $p \in V(T)$ it follows that $\lambda(p)$ is an edge-induced vertex-cut in $H$.*

Theorem 7 shows that there is a close relationship between edge-induced vertex-cuts and hypertree decompositions. Of course, the $\lambda$-sets are not necessarily *minimum* edge-induced vertex-cuts in the underlying hypergraph $H$. However, they may be minimum edge-induced vertex-cuts of the subhypergraphs obtained by splitting $H$ according to the previous edge-induced vertex-cut, or there may always be at least one minimum edge-induced vertex-cut in $H$. We believe that a thorough investigation of this relationship will lead to a better understanding of hypertree decompositions and related techniques.

Finally, let us mention that there exist several powerful heuristic approaches for hypergraph partitioning which compute minimum $s$-$t$-edge-cuts in their final step after determining $s$ and $t$ heuristically. Experimental results have shown that the construction of hypertree decompositions based on such hypergraph partitioning heuristics leads in general to bad results, i.e., the hypertree-width becomes too large. One reason for this may be that hypergraph partitioning heuristics aim at minimizing edge-cuts while hypertree decompositions aim at minimizing edge-induced vertex-cuts. However, the development of efficient approximation algorithms for the computation of minimum edge-induced vertex-cuts, which can then be used as part of hypergraph partitioning heuristics, may improve these results.

## 7  Conclusion

We introduced the notion of edge-induced vertex-cuts which is closely related to the concept of hypertree decomposition. We gave a systematic overview of graph and hypergraph cuts in the literature and compared them to our new edge-induced vertex-cut. We proved that computing minimum $s$-$t$-edge-induced vertex-cuts and minimum edge-induced vertex-cuts is NP-hard; in the case of hypergraphs, we could further show that these problems are fixed-parameter intractable. In future research we will try to determine the fixed-parameter complexity in the case of ordinary graphs. Moreover, the development of efficient approximation algorithms and heuristics for computing minimum edge-induced vertex-cuts would be helpful in several applications.